\newif\ifsingle

\ifsingle
\documentclass[12pt,draftclsnofoot, onecolumn]{IEEEtran}		
\else		
\documentclass[10pt,final, twocolumn]{IEEEtran}
\fi

\usepackage[noadjust]{cite}
\usepackage[T1]{fontenc}
\usepackage[latin9]{inputenc}
\usepackage{color}
\usepackage{amsmath}
\usepackage{amssymb}
\usepackage{graphicx}
\usepackage{textcomp}
\usepackage{multirow}
\usepackage{xcolor}
\usepackage[bookmarks,hidelinks]{hyperref}

\makeatletter

\providecommand{\tabularnewline}{\\}

\ifsingle
 
\else
 
\fi 

\definecolor{NewColor}{rgb}{0.2,0,0.5}

\makeatother

\begin{document}
\title{Machine Learning-Based Intrusion Detection: Feature Selection versus Feature Extraction}
\author{Vu-Duc Ngo, Tuan-Cuong Vuong, Thien Van Luong, and Hung Tran

\thanks{Vu-Duc Ngo is with \mbox{MobiFone} Research and Development Center, \mbox{MobiFone} Corporation, Hanoi 11312, and also with the School of Electronics and Electrical Engineering,
Hanoi University of Science and Technology, Hanoi 11657, Vietnam.
(email: duc.ngo@mobifone.vn).}

\thanks{Thien Van Luong, Tuan-Cuong Vuong, and Hung Tran are with the Faculty of Computer Science, Phenikaa University, Hanoi 12116, Vietnam (e-mail: cngvng123@gmail.com, \{thien.luongvan, hung.tran\}@phenikaa-uni.edu.vn) 
} 
\thanks{Corresponding author: Hung Tran}

\vspace{-0.5cm}
 }
 
\maketitle
\vspace{-1cm}

\begin{abstract}
Internet of things (IoT) has been playing an important role in many sectors, such as smart cities, smart agriculture, smart healthcare, and smart manufacturing. However, IoT devices are highly vulnerable to cyber-attacks, which may result in security breaches and data leakages. To effectively prevent these attacks, a variety of machine learning-based network intrusion detection methods for IoT networks have been developed, which often rely on either feature extraction or feature selection techniques for reducing the dimension of input data before being fed into machine learning models. This aims to make the detection complexity low enough for real-time operations, which is particularly vital in any intrusion detection systems.  \textcolor{black}{This paper provides a comprehensive comparison between these two feature reduction methods of intrusion detection in terms of various performance metrics, namely, precision rate, recall rate, detection accuracy, as well as runtime complexity, in the presence of the modern UNSW-NB15 dataset as well as both binary and multiclass classification. For example, in general, the feature selection method not only provides better detection performance but also lower training and inference time compared to its feature extraction counterpart, especially when the number of reduced features $K$ increases. However, the feature extraction method is much more reliable than its selection counterpart, particularly when $K$ is very small, such as $K=4$. Additionally, feature extraction is less sensitive to changing the number of reduced features $K$ than feature selection, and this holds true for both binary and multiclass classifications. Based on this comparison, we  provide a useful guideline for selecting a suitable intrusion detection type for each specific scenario, as detailed in Tab.~\ref{tab:sum-comparison} at the end of Section~\ref{sec:results}. Note that such the comparison between feature selection and feature extraction over UNSW-NB15 as well as  theoretical guideline have been overlooked in the literature.}
\end{abstract}


\vspace{-0.2cm}

\section{Introduction}

Internet of Things (IoT) has recently witnessed an explosive expansion in a 
 broad range of daily life and industrial applications \cite{al2015internet, Chaabouni2019Network, kumar2022secure}, such as healthcare, smart homes, smart cities, smart energy, smart agriculture, and intelligent transportation. The IoT networks aim to provide internet connections for transferring data among massive IoT devices, such as interconnected sensors, drones, actuators, smart vehicles and smart home appliances \cite{Chaabouni2019Network}, using either wired or wireless communications. However, most of these IoT devices are low-cost, low-power and limited-resource, making them highly vulnerable to cyber attacks as well as intrusive activities. Therefore, it is vital to develop network intrusion detection systems (NIDS) that can promptly and reliably identify and prevent malicious attacks to IoT networks. For this, a wide range of machine learning-based intrusion detection techniques have been designed for IoT, along with a number of public network traffic datasets \cite{Chaabouni2019Network,Mishra2019Detailed}. These datasets often contain a large number of features, in which many are irrelevant or redundant, which adversely affect both the complexity and accuracy of machine learning algorithms. Thus, many feature reduction methods have been developed for NIDS, in which feature selection and feature extraction are two of the most popular ones \cite{Chaabouni2019Network,Mishra2019Detailed}, as discussed next.\footnote{\textcolor{black}{Note that several recent works that apply deep learning and blockchain to secure IoT networks can be found in \cite{kumar2022deep, kumar2023blockchain, Angelo2021effective}, in the fields of healthcare system, unmanned aerial vehicle and Android malware.} }

In NIDS, feature selection has been widely used for reducing the dimensionality of original traffic data. \textcolor{black}{For example, in \cite{ambusaidi2016building}, a mutual information (MI)-based feature selection algorithm was proposed in combination with a classifier called least square support vector machine, which achieves higher accuracy and lower runtime complexity than the existing schemes, over three datasets, namely, KDD99 \cite{KDD99}, NSLKDD  \cite{Mahbod2009KDD} and
Kyoto 2006+ \cite{Song2011Kyoto}. Before that a MI-based scheme was also proposed for NIDS in \cite{amiri2011mutual}, which however suffers from higher computational complexity than the approach in \cite{ambusaidi2016building}. Additionally, several approaches that rely on genetic algorithm (GA) as a search strategy to select the best subset of features can be found in \cite{KHAMMASSI2017GA,Shahri2016Hybrid}. These methods provide lower false alarm rates than the baselines, where UNSW-NB15 \cite{Moustafa2015Dataset} and KDD99 \cite{KDD99} datasets are used. In \cite{Moustafa2017Central}, a hybrid feature selection approach, which relies on the association rule mining and the central points of attribute values, was developed, showing that UNSW-NB15 dataset achieves a better evaluation than NSLKDD. In \cite{Tama2019Two}, another hybrid feature selection method that comprises particle swarm optimization (PSO), ant colony algorithm, and GA was proposed, learning to better detection performance than the baselines such as GA \cite{KHAMMASSI2017GA}, in the presence of both NSLKDD and UNSW-NB15 datasets. In \cite{ALAZZAM2020Pigeon}, a Pigeon inspired optimizer was used for selecting features of NIDS, which achieves higher accuracy than the PSO \cite{KHAMMASSI2017GA} and hybrid association rules methods \cite{Moustafa2017Central}. Note that the aforementioned feature selection schemes often suffer from high computational cost, especially for those relying on GA, PSO or machine learning-based classifiers. For this, a correlation-based feature selection method that offers low computational cost was investigated for NIDS over KDD99 and UNSW-NB15 datasets in \cite{Moustafa2016Evaluation}, taking the correlation level among features into account. Recently, this correlation-based method was combined with ensemble-based machine learning classifiers to significantly improve the accuracy of NIDS \cite{Moustafa2019Ensemble}, at the cost of higher complexity.} Hence, aiming at a real-time and low-latency attack detection solutions, this work will more focus on the correlation-based feature selection method.\footnote{\textcolor{black}{Note that several matrix factorization-based dimensionality reduction methods were developed for gene expression analysis in \cite{saberi2022dual, azadifar2022graph}}.}

Unlike feature selection, which retains a subset of the original features in NIDS, feature extraction attempts to compress a large amount of original features into a low-dimensional vector so that most of information is retained.  There are a number of feature extraction techniques that have been applied for reducing data dimension in NIDS, such as principal component analysis (PCA), linear discriminant analysis (LDA), and neural network-based autoencoder (AE). \textcolor{black}{For instance, in \cite{Xu2005PCA}, PCA was applied to significantly reduce the dimension of KDD99 dataset, improving both accuracy and speed of NIDS, where support vector machine was used for attack classification. Then, several variants of PCA were adopted to intrusion detection, such hierarchical PCA neural networks \cite{liu2007hierarchical} and kernel PCA with GA \cite{kuang2014novel}, which can enhance the detection precision for low-frequent attacks. Some of applications of PCA to recent network traffic datasets such as UNSW-NB15 and CICIDS2017 \cite{Iman2018Toward} can be found in \cite{Razan2019PCA, Qi2022Fast}. In addition to PCA, LDA was also employed as a feature reduction method for NIDS in \cite{Tan2010LDA}, which remarkably reduces the computational complexity of NIDS. Then, in \cite{pajouh2019two,Pajouh2017Two}, both PCA and LDA were combined into a two-layer dimension reduction, which is capable of reliably detecting low-frequency malicious activities, such as User to Root and Remote to Local, over NSLKDD dataset. To further improve the efficiency of feature extraction in NIDS, a AE-based neural network was used in a range of research works \cite{yan2018effective,Khan2019Novel, Popoola2021Hybrid, Zhou2021LSTM, dao2021stacked, d2021network}. In particular, a stacked sparse AE approach was developed in \cite{yan2018effective} to conduct a non-linear mapping between high-dimensional data and low-dimensional data over NSLKDD dataset. In \cite{Khan2019Novel}, a deep stacked AE was used to noticeably reduce the number of features to 5 and 10 for binary and multiclass classification, respectively, leading to better accuracy than the previous methods. Additionally, a number of AE architectures based on long short-term memory (LSTM) were developed for dimensionality reduction of NIDS, such as variational LSTM \cite{Zhou2021LSTM} and bidirectional LSTM \cite{Popoola2021Hybrid}, which can  efficiently address imbalance and high-dimensional problems. Note that these AE-based methods suffer from a high computational cost compared to PCA and LDA, both in training and testing phases. To address this issue, a network pruning algorithm has been recently proposed in  
\cite{dao2021stacked} to considerably lower complexity of AE structures in extracting features of NIDS.} \textcolor{black}{In \cite{d2021network}, a network architecture  uses an autoencoder based on convolutional and recurrent neural networks to extract spatial and temporal features without human engineering.}

\textcolor{black}{It is worth noting that most of the aforementioned papers have focused on either improving the detection accuracy or reducing the computational complexity of NIDS, by using machine learning-based classifications in combination with feature engineering methods such as feature selection and feature extraction for reducing data dimensionality. However, a comprehensive comparison between these two feature reduction methods has been overlooked in the literature. Our paper appears to address this gap. In particular, we first provide an overview of NIDS, with a focus on the phase of feature reduction, where feature extraction with PCA and feature selection with correlation matrix are the two promising candidates for realistic low-latency operations of NIDS. Then, using the modern UNSW-NB15 dataset, we thoroughly compare the detection performance (precision, recall, F1-score) as well as runtime complexity (training time and inference time) of these two methods, taking into account both binary and multiclass classifications as well as the same number of selected/extracted features denoted as $K$. Based on our extensive experiments, we found that feature selection generally achieves higher detection accuracy and requires less training and inference times when the number of reduced features $K$ is large enough, while feature extraction outperforms feature selection when $K$ gets smaller, such as $K=4$ or less. Furthermore, in order to gain a deeper insight into detection behaviors of both methods, we investigate and compare their accuracy for each attack class when varying $K$, based on their best machine learning classifiers, which revealed that feature extraction is not only less sensitive to varying the number of reduced features but also capable of detecting more diverse attack types than feature selection. Additionally, both tend to be able to detect more attacks, i.e., Abnormal classes, when having more features selected or extracted. Relying on such comprehensive observations, we provide a theoretical guideline for selecting an appropriate intrusion detection type for each specific scenario, as detailed in Tab.~\ref{tab:sum-comparison} at the end of Section~\ref{sec:results}, which is, to the best of our knowledge, not available in the literature.}

The rest of this paper is organized as follows. Section II discusses machine learning-based network intrusion detection methods for IoT networks. The overview of UNSW-NB15 dataset and data pre-processing are explained in Section III. Section IV provides the experimental results and discussion. Finally, Section V concludes this paper.

\vspace{-0.2cm}

\section{Machine Learning-based Network Intrusion Detection Methods \label{sec:idiot}}

In this section, we describe an overview of a network intrusion detection system (NIDS) based on machine learning, followed by details on the two major feature reduction methods, namely, feature selection and feature extraction.

\vspace{-0.2cm}

\subsection{Overview of NIDS\label{subsec:nids}}

\vspace{-0.1cm}
 \begin{figure*}[ht]
\begin{centering}
\includegraphics[width=1.8\columnwidth]{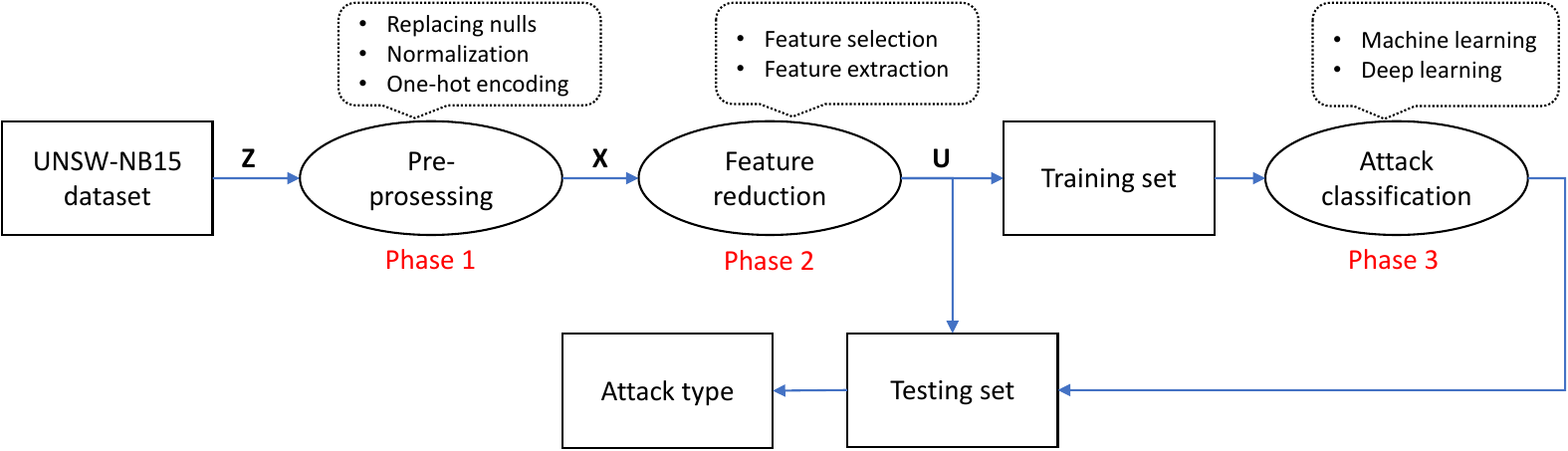}
\par
\centering\caption{Block diagram of a network intrusion detection system. \label{fig:nids}}
\end{centering}
\end{figure*}
\textcolor{black}{A NIDS consists of three major components, namely, data pre-processing, feature reduction, and attack classification, as illustrated in Fig.~\ref{fig:nids}. In particular, in the first phase, the raw data is denoted as the dataframe
 $\mathbf{Z}$, whose features may include unexpected or non-numeric values, such as null or nominal. $\mathbf{Z}$ is pre-processed in order to either replace these unexpected values with valid ones or transform them to the numeric format using one-hot encoding. Several features that do not affect detection performance, such as the source IP address and the source port number, are dropped out. Furthermore, depending on the classifier we use for identifying attacks, we may use the normalization technique, for example, to constrain the values of all features, i.e., the elements of the output vector of the first phase $\mathbf{X}$ in Fig.~\ref{fig:nids}, to range from 0 to 1. We will discuss this in detail in Section~\ref{sec:dataset} when presenting UNSW-NB15 dataset.}

 \textcolor{black}{As such, after the first phase, the pre-processed data $\mathbf{X}\in\mathbb{R}^{D\times N}$ is likely to have much more features than the original data $\mathbf{Z}$, particularly due to the use of one-hot encoding, where $D$ is the number of dimensions, or equivalently, the number of features of $\mathbf{X}$, and $N$ is the number of data samples. For example, when UNSW-NB15 dataset is used, the dimension of data increases from 45 to nearly 200, which is too large for classification techniques to quickly recognize the attack type. In order to address this fundamental issue, in the second phase, we need to reduce the number of features that will be used for the attack classification phase (the last phase in Fig.~\ref{fig:nids}). For this, two feature reduction methods called feature selection and feature extraction are widely used to either select or extract a small number of most important features from pre-processed traffic data. This procedure also helps to remove a large amount of unnecessary features, which not also increase the complexity of NIDS, but also degrade its detection performance, as will be illustrated in experimental results in Section~\ref{sec:results}. Herein, the output data of the feature reduction block is denoted as vector $\mathbf{U}\in\mathbb{R}^{K\times N}$ in Fig.~\ref{fig:nids}, which is expected to have a much lower dimension than $\mathbf{X}$, i.e., $K\ll D$, while retaining its most important information. }
 
\textcolor{black}{Finally, in the third phase of NIDS, a number of binary and multiclass classification approaches based on machine learning, such as decision tree,  random forest and multilayer perception neural networks, are employed to detect the attack type. Relying on attack detection results, the system administrators can promptly make a decision to prevent malicious activities, ensuring the security of IoT networks. Here, note that the detection performance and latency of a NIDS strongly depend on which classifier and  which feature reduction method it employs. Therefore, in this contribution, we comprehensively investigate detection performance (in terms of recall, precision, F1-score) and latency (in terms of training time and inference time) of different detection methods in presence of both feature selection and feature extraction as well as different machine learning classifiers. We also focus more on the comparison between these two feature reduction methods, which will be described in detail in the following subsections.}

\vspace{-0.2cm}

\subsection{Feature selection\label{subsec:fs}}
\vspace{-0.1cm}
 There are a number of feature selection techniques used in intrusion detection, namely, information gain (IG) \cite{ambusaidi2016building} and feature correlation \cite{hall1999correlation,Moustafa2019Ensemble,Moustafa2016Evaluation}. In this work, we focus on using feature correlation for selecting important features, since this method has been shown to achieve competitive detection accuracy and complexity compared to other selection counterparts. \textcolor{black}{Using this correlation-based method, we aim to select features that are most correlated to other features based on the correlation matrix calculated from the training dataset. More specifically, the correlation coefficient between feature $\Omega_1$ and feature  $\Omega_2$ is calculated based on the numeric pre-processed training dataset $\mathbf{X}$ as follows \cite{hall1999correlation}:
 \begin{equation}
\mathcal{C}_{\Omega_{1},\Omega_{2}}=\frac{\sum_{i=1}^{N}\left(\alpha_{i}-E_{\Omega_{1}}\right)\left(\beta_{i}-E_{\Omega_{2}}\right)}{\sqrt{\sum_{i=1}^{N}\left(\alpha_{i}-E_{\Omega_{1}}\right)^{2}}.\sqrt{\sum_{i=1}^{N}\left(\beta_{i}-E_{\Omega_{2}}\right)^{2}}},\label{eq:corr_coeif}
 \end{equation}
where $\alpha_i$ and $\beta_i$ are the values of these two features, $E_{\Omega_{1}}=\sum_{i=1}^{N}\alpha_{i}/N$ and $E_{\Omega_{2}}=\sum_{i=1}^{N}\beta_{i}/N$ are their means over $N$ training data samples. Note that after preprocessing the raw data $\mathbf{Z}$ to obtain $\mathbf{X}$, all features of $\mathbf{X}$ are now numeric, i.e.,  $\alpha_i$ and $\beta_i$ are numeric, making \eqref{eq:corr_coeif} applicable to process.}  By doing this, we obtain a $D\times D$ correlation matrix $\mathbf{C}$, whose elements are given by $c_{ij}=\mathcal{C}_{\Omega_{i},\Omega_{j}}$ for $i,j=1,2,...,D$. The average correlation of feature $\Omega_i$ to other features is computed as follows: 
\begin{equation}
C_i=\frac{\sum_{j=1}^{D}c_{ij}}{D},\label{eq:Ci}
\end{equation}
where $c_{ii}=1$ for $j=i$ and $c_{ij}\in [-1;1]$ for $j\neq i$. Note that the self-correlation coefficient $c_{ii}$ does not affect selection results, since it contributes the same amount to all $C_i$ for $i=1,2,...,D$. Then, using a suitable threshold, as will be detailed in Section~\ref{sec:results}, we are able to select $K$ most important features corresponding to $K$ largest elements $C_i$.

It is worth noting that we only need to calculate such feature correlation in the training phase, while in the testing phase, we simply pick up $K$ features from the high-dimensional data $\mathbf{X}$ to form the reduced-dimensional data $\mathbf{U}$ in Fig.~\ref{fig:nids}. This does not require much computational resource when compared with the feature extraction method, which is presented next.

\vspace{-0.2cm}

\vspace{-0.2cm}

\subsection{Feature extraction\label{subsec:fe}}

\vspace{-0.1cm}
 Principal component analysis (PCA) \cite{Xu2005PCA} and autoencoder (AE) \cite{dao2021stacked} are the two major feature extraction methods used in the NIDS. Different from feature selection, whose selected features are identical to those appearing in the original data, these feature extraction techniques compress the high-dimensional data $\mathbf{X}$ into the low-dimensional data $\mathbf{U}$ using either a projection matrix or an AE-based neural network learned from training dataset. Note that the AE approach usually suffers from high computational complexity of a deep neural network (DNN), leading to higher latency than the PCA. Thus, in this work, we concentrate on the PCA-based feature extraction approach in order to fulfill a strict requirement on the latency of the NIDS for promptly preventing severe cyber attacks. 

In what follows, we introduce the procedure of producing the $D\times K$ projection matrix $\mathbf{W}$ in the training phase, and how to utilize this matrix in the testing phase. In particular, based on the pre-processed training data $\mathbf{X}$ of $N$ samples, we normalize it by subtracting all samples of $\mathbf{X}$ by its mean over all training samples, i.e., the normalized data is given as follows: $\hat{\mathbf{X}}=\mathbf{X}-\bar{\mathbf{X}}$, where $\bar{\mathbf{X}}$ is the mean vector. Then, we compute the $D\times D$ covariance matrix of training data as follows: $\mathbf{R}=\frac{1}{N}\hat{\mathbf{X}}\hat{\mathbf{X}}^{T}$. Based on this, we determine its eigenvalues and eigenvectors, from which, we select $K$ eigenvectors corresponding to $K$ largest eigenvalues for constructing the $D\times K$ projection matrix $\mathbf{W}$. Herein, these $K$ eigenvectors are regarded as the principal components that create a subspace, which is expected to be significantly close to the normalized high-dimensional data $\hat{\mathbf{X}}$. Finally, the compressed data is determined by $\mathbf{U}=\mathbf{W}^{T}\hat{\mathbf{X}}$, which now has the size of $K\times N$ instead of $D\times N$ of the original data.

In the testing phase, for each new data point $\mathbf{x}_i\in\mathbb{R}^D$, its dimension is  reduced using PCA according to $\mathbf{u}_{i}=\mathbf{W}^{T}\left(\mathbf{x}_{i}-\bar{\mathbf{X}}\right)$. This indicates that the output of the training phase of PCA includes both the projection matrix $\mathbf{W}$ and the mean vector of all training samples $\bar{\mathbf{X}}$. It should be noted that such projection matrix calculation would be computationally expensive, particularly when $D$ and $K$ are large.


\section{Overview of UNSW-NB15 dataset\label{sec:dataset}}
We now present some key information about UNSW-NB15 dataset, which will be used in our experiments in Section~\ref{sec:results} to compare between feature selection and feature extraction. Then, the data pre-processing for this dataset is also discussed.

\begin{figure}[t]
\begin{centering}
\includegraphics[width=1\columnwidth]{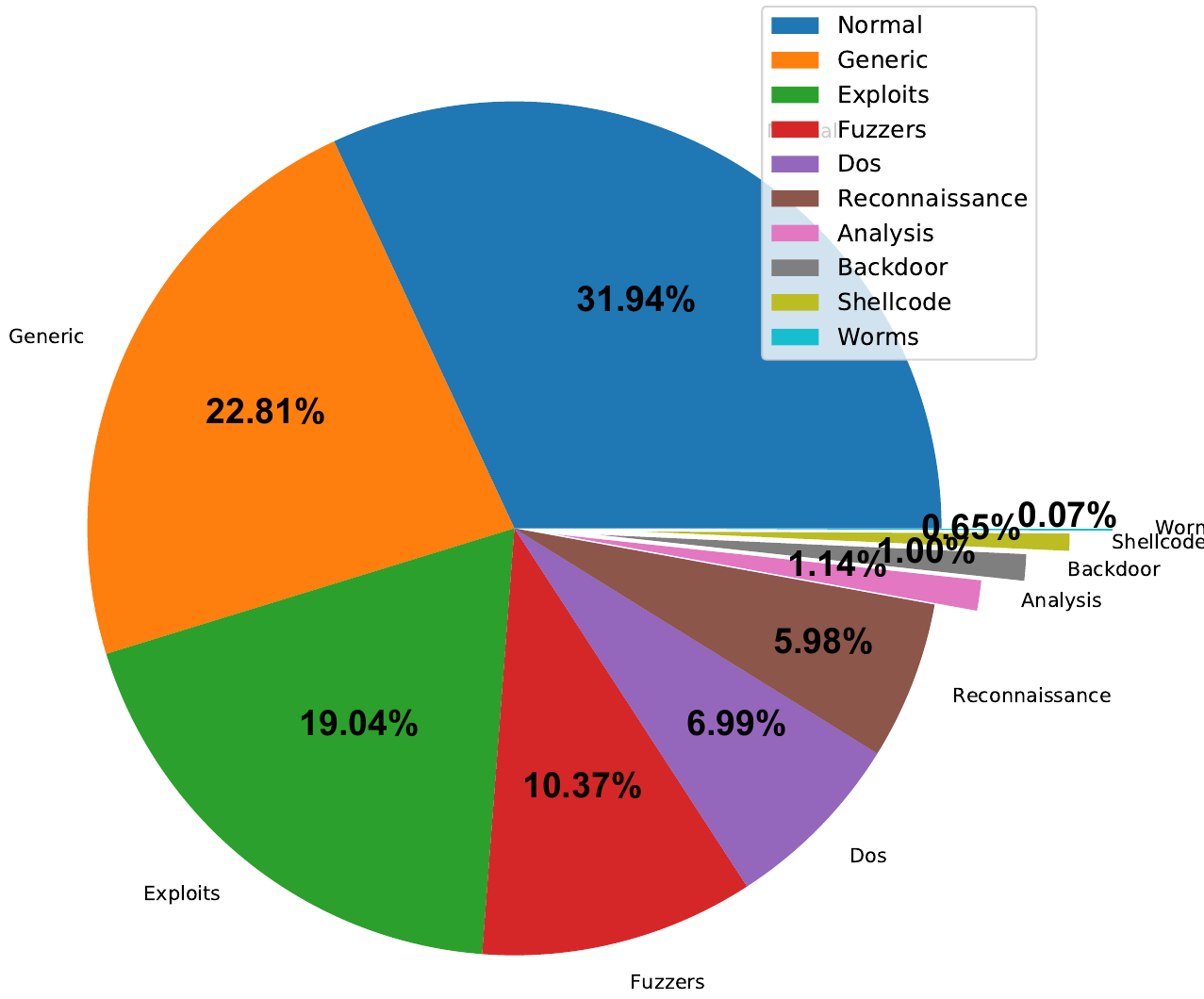}
\par\end{centering}
\caption{Proportions of 10 classes in training dataset of UNSW-NB15. \label{fig:Pc-tr}}
\end{figure}

\subsection{Key information of UNSW-NB15 dataset}\label{subsec:dataset}

\begin{figure}[t]
\begin{centering}
\includegraphics[width=1\columnwidth]{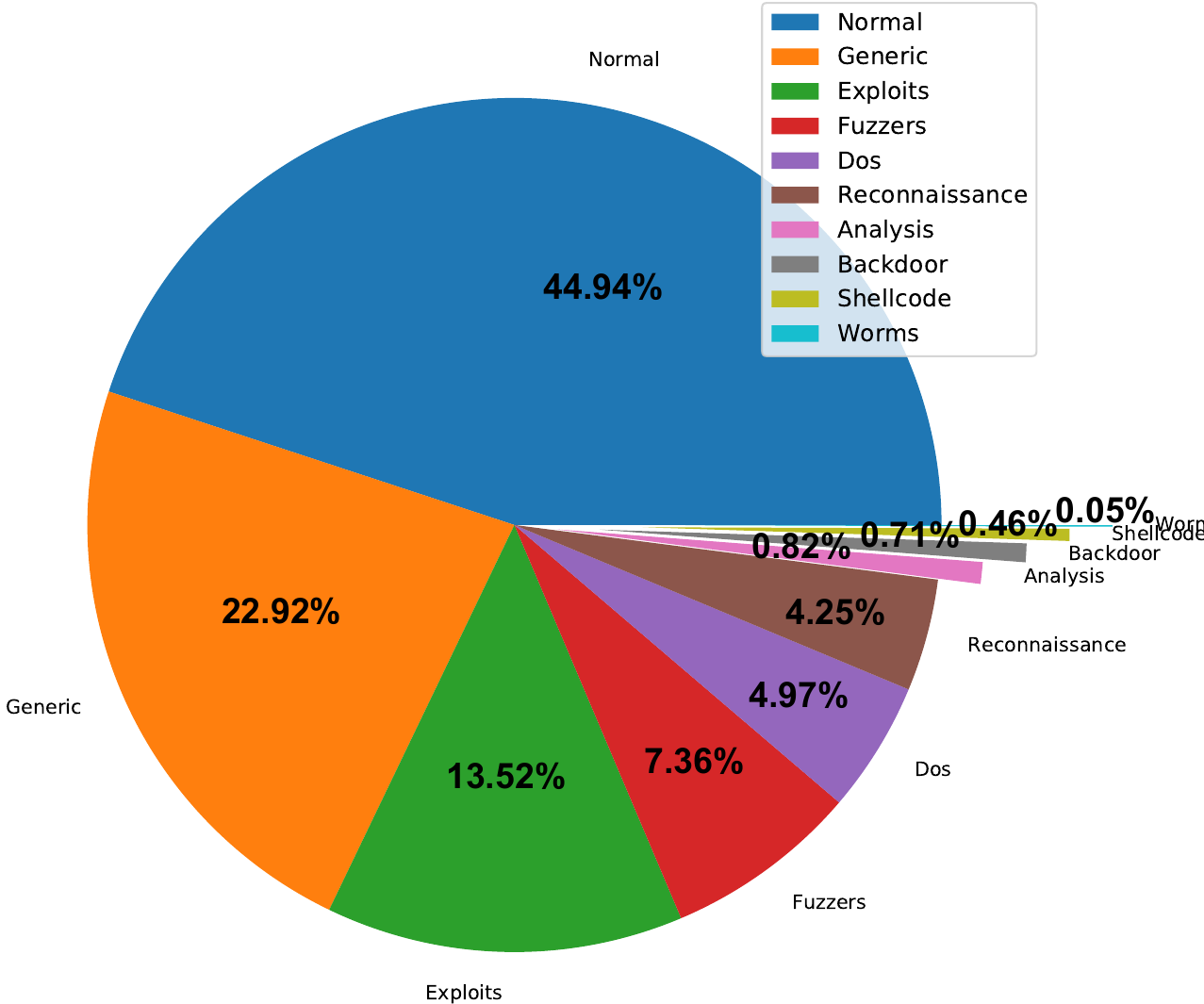}
\par\end{centering}
\caption{Proportions of 10 classes in testing dataset of UNSW-NB15. \label{fig:Pc-test}}
\end{figure}

UNSW-NB15 dataset was first introduced in \cite{Moustafa2015Dataset}, which offers better real modern normal and abnormal synthetical network traffic compared with the previous NIDS datasets such as KDD99 \cite{KDD99} and NSLKDD \cite{Mahbod2009KDD}. A total of 2.5 million records of data are included in the UNSW-NB15 dataset, in which there are one normal class and nine attack classes: Analysis, Backdoor, DoS, Exploits, Fuzzers, Generic, Reconnaissance, Shellcode, and Worms. Flow features, basic features, content features, time features, additional generated features, and labeled features are six feature groups, which consist of a total of 49 features in the original data \cite{Moustafa2015Dataset}. However, in this work, we use a 10\% cleaned dataset of UNSW-NB15, which includes a training set of 175,341 records and a test set of 82,332 records. There are a few minority classes with proportions of less than 2\%, including Analysis, Backdoor, Shellcode, and Worms (see Fig.~\ref{fig:Pc-tr} and Fig.~\ref{fig:Pc-test}). In the 10\% dataset, some unrelevant features were removed, such as \textit{scrip} (source IP address), \textit{sport} (source port number), \textit{dstip} (destination IP address), and \textit{dsport} (destination port number). Therefore, the number of features was reduced to 45, including 41 numerical features and 4 nominal features.

\subsection{Pre-processing dataset}\label{subsec:prepreocess}
As mentioned above, the 10\% dataset of UNSW-NB15 has 45 features, including 41 numerical features and 4 nominal features. We remove the \emph{id} feature in numerical features, since it does not affect the detection performance. The \emph{attack\_cat} nominal feature that contains the names of attack categories is also removed. Thus, there are 3 remaining helpful nominal features, namely, \textit{proto, service, state}. In addition, null values appearing in the \emph{service} feature are treated as 'other' type of service. 

\begin{table}[]
\caption{An example of one-hot encoding for a nominal feature}
\centering
\begin{tabular}{|c|c|c|c|}
\hline 
\emph{proto} & \emph{proto\_A} & \emph{proto\_B} & \emph{proto\_C}\tabularnewline
\hline 
\hline 
A & 1 & 0 & 0\tabularnewline
\hline 
B & 0 & 1 & 0\tabularnewline
\hline 
C & 0 & 0 & 1\tabularnewline
\hline 
\end{tabular}
\label{tab:one-hot}
\end{table}

  One-hot encoding is used for transforming nominal features, i.e., \textit{proto, service, state}, to numerical values. For example, assume that the \textit{proto} feature has a total of 3 different values, namely, A, B, C, then its one-encoding will result in 3 numerical features, namely,  \textit{proto\_A, proto\_C, proto\_C}, whose values are 0 or 1, as illustrated in Tab.~\ref{tab:one-hot}.  As a result, after pre-processing data, the number of features will increase from 45 features in \textbf{Z} to approximately 200 features in \textbf{U} (see Fig.~\ref{fig:nids}), where many of them are not really helpful in classifying attacks. Therefore, it is necessary to reduce such a large number of features to a few of the most important features, which allows to reduce the complexity of machine learning models in the classification phase. Finally, we note that when feature extraction is used, we normalize the input feature with the min-max normalization method \cite{minmaxscalaer} to improve the classification accuracy, while we do not use that data normalization for feature selection, since it does not improve the performance.

\vspace{-0.2cm}
\section{Experimental results and discussion\label{sec:results}}

 We now present extensive experimental results for investigating the performance of the NIDS using both feature selection and feature extraction methods described in Section~\ref{sec:idiot}, in combination with a range of machine learning-based classification models. More particularly, the performance metrics used for comparison include recall (R), precision (P), F1-score, training time and inference time, which will be explained in detail in Subsection~\ref{subsec:setting}. Both binary and multiclass classifications are considered. We also investigate the accuracy for each attack class to provide an insight into the behaviors of different detection methods. Last but not least, based on our extensive comparison between feature selection and feature extraction, we provide a helpful guideline on how to choose an appropriate detection technique for each specific scenario. 


\subsection{Implementation setting\label{subsec:setting}}

\subsubsection{Computer configuration}\label{subsubsub:computer}
The configuration of our computer, its operation system as well as  a range of Python packages used for implementing intrusion detection algorithms in this work are detailed in Tab.~\ref{tab:setting}. 

\begin{table}[]    \caption{Hardware and Environment Specification}
    \centering
    \begin{tabular}{r|l}
    \hline 
    \textbf{Unit} & \textbf{Description}\tabularnewline
    \hline 
    Processor & Intel Core i5-10400F (2.66 Hz, \\
              & 6 cores 12 threats, 12MB Cache, 65W)\tabularnewline
    \hline 
    RAM & 16GB \tabularnewline
    \hline 
    GPU & Nvidia GTX 1650 OC-4G\tabularnewline
    \hline 
    Operating System & Ubuntu 20.04.4 LTS\tabularnewline
    \hline 
    Packages & Numpy, Matplotlib, Pandas, Scipy, \\
             &  Scikit-learn, Scikit-plot and Time\tabularnewline
    \hline 
    \end{tabular}

    \label{tab:setting}
\end{table}

\begin{table}[]

\caption{Threshold setting and Features selected}
    \centering
    \begin{tabular}{r|c|l}
    \hline 
    \textbf{Threshold} & \textbf{Number} & \textbf{Features selected}\tabularnewline
    \hline 
    0.011 & 4 & 'spkts', 'dpkts', 'dbytes', 'dloss'\tabularnewline
    \hline 
    0.0137 & 8 & 'dur', 'spkts', 'dpkts', 'sbytes', 'dbytes', \\
        & & 'sloss', 'dloss','ct\_state\_ttl'\tabularnewline
    \hline 
    0.011 & 16 & 'dur', 'spkts', 'dpkts', 'sbytes', 'dbytes', 'sloss', \\
    & &    'dloss',       'dinpkt', 'sjit',  'djit', 'tcprtt', 'synack', \\
    & &   'ackdat',        'response\_body\_len', 'ct\_state\_ttl',\\
    & &  'proto\_icmp'\tabularnewline
    \hline 
    \end{tabular}\label{tab:fsthreshold}

\end{table}

\begin{table*}[ht]
\centering \caption{Feature Selection versus Feature Extraction: 4 selected/extracted features and binary classification}
\label{tab:binary-4} \linespread{1.0} 
\begin{tabular}{|l|c|c|c|c|c|c|c|c|c|c|}
\hline 
\multirow{2}{*}{Models} & \multicolumn{5}{c|}{Feature Extraction} & \multicolumn{5}{c|}{Feature Selection}\tabularnewline
\cline{2-11} \cline{3-11} \cline{4-11} \cline{5-11} \cline{6-11} \cline{7-11} \cline{8-11} \cline{9-11} \cline{10-11} \cline{11-11} 
 & P & R & F1 & training (s) & inference ($\mu$s) & P & R & F1 & training (s) & inference ($\mu$s)\tabularnewline
\hline 
Decision Tree & 85.33 & 84.35 & 84.84 & 22.73 & 3.73 & \textbf{84.09} & \textbf{79.89} & \textbf{81.94} & 14.53 & \textcolor{red}{\textbf{0.07}}\tabularnewline
\hline 
Random Forest & 85.76 & 81.22 & 83.42 & 32.32 & 18.12 & 77.86 & 75.11 & 76.46 & 17.13 & 3.29\tabularnewline
\hline 
KNeighbors & \textcolor{red}{\textbf{86.19}} & \textcolor{red}{\textbf{84.67}} & 
\textcolor{red}{\textbf{85.42}}& 23.04 & 38.47 & 52.38 & 47.98 & 50.08 & 14.75 & 259.03\tabularnewline
\hline 
MLP & 85.91 & 81.75 & 83.78 & 1011.31 & 39.37 & 75.76 & 74.75 & 75.25 & 1278.61 & 37.08\tabularnewline
\hline 
Naive Bayes & 72.62 & 71.95 & 72.28 & \textbf{20.37} & \textbf{3.62} & 75.47 & 73.63 & 74.54 & \textcolor{red}{\textbf{14.3}} & 0.24\tabularnewline
\hline 
\end{tabular}
\end{table*}

\begin{table*}[ht]
\centering \caption{Feature Selection versus Feature Extraction: 8 selected/extracted features and binary classification}
\label{tab:binary-8} \linespread{1.0} 
\begin{tabular}{|l|c|c|c|c|c|c|c|c|c|c|}
\hline 
\multirow{2}{*}{Models} & \multicolumn{5}{c|}{Feature Extraction} & \multicolumn{5}{c|}{Feature Selection}\tabularnewline
\cline{2-11} \cline{3-11} \cline{4-11} \cline{5-11} \cline{6-11} \cline{7-11} \cline{8-11} \cline{9-11} \cline{10-11} \cline{11-11} 
 & P & R & F1 & training (s) & inference ($\mu$s) & P & R & F1 & training (s) & inference ($\mu$s)\tabularnewline
\hline 
Decision Tree & 85.98 & \text{84.98} & 85.48 & 26.94 & \textbf{3.95} & \textcolor{red}{\textbf{87.87}} & \textcolor{red}{\textbf{87.07}} & \textcolor{red}{\textbf{87.47}} & 14.57 & \textcolor{red}{\textbf{0.11}}\tabularnewline
\hline 
Random Forest & 85.23 & 80.03 & 82.55 & 33.89 & 18.32 & 85.74 & 80.95 & 83.27 & 18.08 & 3.34\tabularnewline
\hline 
KNeighbors & \textbf{86.39} & \textbf{84.99} & \textbf{85.69} & 23.41 & 44.02 & 87.08 & 85.98 & 86.53 & 14.99 & 51.16\tabularnewline
\hline 
MLP & 86.14 & 82.42 & 84.24 & 1252.49 & 39.82 & 84.42 & 79.95 & 82.12 & 607.66 & 33.06\tabularnewline
\hline 
Naive Bayes & 72.14 & 70.01 & 71.06 & \textbf{21.76} & 4.11 & 75.56 & 73.85 & 74.69 & \textcolor{red}{\textbf{14.33}} & 0.35\tabularnewline
\hline 
\end{tabular}
\end{table*}

\subsubsection{Evaluation Metrics\label{subsec:metrics}}
We consider the following performance metrics: precision, recall, F1-score, as well as training time and inference time. In particularly, F1-score is calculated based on precision and recall as follows:

\begin{equation}
\text{F1-score}=2\times\frac{\text{precision\ensuremath{\,\times\,\text{recall}}}}{\text{precision}+\text{recall}}, \label{eq:f1score}
\end{equation}
which is regarded as a harmonic mean of precision and recall.

As shown in Fig.~\ref{fig:nids}, the two feature reduction methods considered in this work go through the same pre-processing data step, so we do not take the time requried for this step into account when estimating their training and inference time. Particularly, the training time consists of the training time of classification models and the time duration consumed by feature reduction in training (FR\_train), as follows:
\begin{equation}
    {\text{training time} = \text{time}_{{train}} + \text{time}_{{FR\_train}}}, \label{eq:traintime}
\end{equation}
Meanwhile, the inference time consists of the prediction time of machine learning classifiers and the time duration required for feature reduction in the testing phase, given by
\begin{equation}
    \text{inference time} = \text{time}_{{predict}} + \text{time}_{FR\_test}. \label{eq:testtime}
\end{equation}
\subsubsection{Classification  models}\label{subsubsec: model sett}

\begin{table*}[ht]
\centering \caption{Feature Selection versus Feature Extraction: 16 selected/extracted features and binary classification}
\label{tab:binary-16} \linespread{1.0} 
\begin{tabular}{|l|c|c|c|c|c|c|c|c|c|c|}
\hline 
\multirow{2}{*}{Models} & \multicolumn{5}{c|}{Feature Extraction} & \multicolumn{5}{c|}{Feature Selection}\tabularnewline
\cline{2-11} \cline{3-11} \cline{4-11} \cline{5-11} \cline{6-11} \cline{7-11} \cline{8-11} \cline{9-11} \cline{10-11} \cline{11-11} 
 & P & R & F1 & training (s) & inference ($\mu$s) & P & R & F1 & training (s) & inference ($\mu$s)\tabularnewline
\hline 
Decision Tree & \textbf{86.43} & \textbf{85.47} & \textbf{85.95} & 37.38 & \textbf{5.38} & \textcolor{red}{\textbf{87.41}} & \textcolor{red}{\textbf{86.61}} & \textcolor{red}{\textbf{87.01}} & 15.18 & \textcolor{red}{\textbf{0.17}}\tabularnewline
\hline 
Random Forest & 85.85 & 81.08 & 83.4 & 58.07 & 19.26 & 85.85 & 81.08 & 83.39 & 52.92 & 19.01\tabularnewline
\hline 
KNeighbors & 86.09 & 84.5 & 85.29 & 38.74 & 1421.15 & 86.09 & 84.5 & 85.29 & 37.68 & 1426.22\tabularnewline
\hline 
MLP & 86.36 & 84.67 & 83.04 & 1344.91 & 31.59 & 81.75 & 74.27 & 77.83 & 661.67 & 40.64\tabularnewline
\hline 
Naive Bayes & 78.2 & 75.59 & 76.87 & \textbf{36.64} & 5.82 & 75.66 & 73.88 & 74.76 & \textcolor{red}{\textbf{14.46}} & 0.55\tabularnewline
\hline 
\end{tabular}
\end{table*}

We use five machine learning models to do both binary and multiclass classification tasks, which are available in Python Scikit-learn library, namely, Decision Tree, Random Forest (max\_depth = 5), K-nearest Neighbors (n\_neighbors = 5), Multi-layer Perceptron (MLP) (max\_iter = 100, hidden\_layer\_sizes = 200), and Bernoulli Naive Bayes. Additionally, for a better insight of feature selection, we provide lists of 4, 8 and 16 selected features in Tab.~\ref{tab:fsthreshold}, as well as the corresponding thresholds of the average correlation used to achieve those numbers of selected features.

\vspace{-0.2cm}

\subsection{Binary classification}

\label{subsec:binary} \vspace{-0.1cm}
We first investigate the detection performance and runtime of  feature selection and feature extraction methods when using binary classification in Tabs.~\ref{tab:binary-4}, \ref{tab:binary-8} and \ref{tab:binary-16} for  4, 8, 16 selected/extracted features, respectively. In these tables, the best values (i.e. the maximum values of precision, recall, and F1-score, and the minimum values of training and inference times at each column of the tables) are highlighted in bold, especially the best values for both feature selection and feature extraction are highlighted both in bold and red color. The training time is measured in second (s), while the inference time for each data sample is measured in millisecond ($\mu$s).

\begin{table*}[ht]
\centering \caption{Accuracy comparison for each class between feature selection and feature extraction using binary classification}
\label{tab:acc-binary-best} \linespread{1.0} 
\begin{tabular}{|c|c|c|c|c|c|c|}
\hline 
\multirow{2}{*}{Class} & \multicolumn{3}{c|}{Feature Extraction (MLP/KNeighbors)} & \multicolumn{3}{c|}{Feature Selection (Decision Tree)}\tabularnewline
\cline{2-7} \cline{3-7} \cline{4-7} \cline{5-7} \cline{6-7} \cline{7-7} 
 & $K=4$  & $K=8$  & $K=16$  & $K=4$ & $K=8$  & $K=16$ \tabularnewline
\hline 
Normal & 70.64 & 71.55 & \textbf{73.79} & 57.21 & \textcolor{red}{\textbf{76.67}} & 76.09\tabularnewline
\hline 
Abnormal & \textbf{96.14} & 96.02 & 94.99 & \textcolor{red}{\textbf{98.4}} & 95.55 & 95.21\tabularnewline
\hline 
Average & 84.68 & 85.02 & \textbf{85.47} & 79.89 & \textcolor{red}{\textbf{87.07}} & 86.61\tabularnewline
\hline 
\end{tabular}

\end{table*}

\begin{table*}[ht]
\centering \caption{Feature Selection versus Feature Extraction: 4 selected/extracted features and multiclass classification}
\label{tab:multiclass-4} \linespread{1.0} %


\begin{tabular}{|l|c|c|c|c|c|c|c|c|c|c|}
\hline 
\multirow{2}{*}{Models} & \multicolumn{5}{c|}{Feature Extraction} & \multicolumn{5}{c|}{Feature Selection}\tabularnewline
\cline{2-11} \cline{3-11} \cline{4-11} \cline{5-11} \cline{6-11} \cline{7-11} \cline{8-11} \cline{9-11} \cline{10-11} \cline{11-11} 
 & P & R & F1 & training (s) & inference ($\mu$s) & P & R & F1 & training (s) & inference ($\mu$s)\tabularnewline
\hline 
Decision Tree & 75.54 & 67.56 & 71.33 & 21.76 & 4.47 & 
\textbf{69.71}& \textbf{61.65} & \textbf{65.43} & 14.45 & \textcolor{red}{\textbf{0.09}}\tabularnewline
\hline 
Random Forest & 77.6 & 64.35 & 70.36 & 30.52 & 20.69 & 55.73 & 58.76 & 57.21 & 17.55 & 5.55\tabularnewline
\hline 
KNeighbors  & 76.98 & \textcolor{red}{\textbf{70.87}} & 73.8 & 20.49 & 38.57 & 50.28 & 45.87 & 47.97 & 14.67 & 258.84\tabularnewline
\hline 
MLP & \textcolor{red}{\textbf{80.1}} & 68.96 & \textcolor{red}{\textbf{74.11}} & 1085.63 & 42.89 & 63.42 & 55.41 & 59.15 & 1519.35 & 43.67\tabularnewline
\hline 
Naive Bayes & 62.82 & 50.81 & 56.18 & \textbf{19.9} & \textbf{3.83} & 41.55 & 59.68 & 48.99 & \textcolor{red}{\textbf{12.29}} & 0.59\tabularnewline
\hline 
\end{tabular}

\end{table*}

\begin{table*}[ht]
\centering \caption{Feature Selection versus Feature Extraction: 8 selected/extracted features and multiclass classification}
\label{tab:multiclass-8} \linespread{1.0} %


\begin{tabular}{|l|c|c|c|c|c|c|c|c|c|c|}
\hline 
\multirow{2}{*}{Models} & \multicolumn{5}{c|}{Feature Extraction} & \multicolumn{5}{c|}{Feature Selection}\tabularnewline
\cline{2-11} \cline{3-11} \cline{4-11} \cline{5-11} \cline{6-11} \cline{7-11} \cline{8-11} \cline{9-11} \cline{10-11} \cline{11-11} 
 & P & R & F1 & training (s) & inference ($\mu$s) & P & R & F1 & training (s) & inference ($\mu$s)\tabularnewline
\hline 
Decision Tree & 76.43 & 69.36 & 72.72 & 27.49 & \textbf{4.42} & 80.18 & \textcolor{red}{\textbf{76.62}} & \textcolor{red}{\textbf{78.36}} & 14.65 & \textcolor{red}{\textbf{0.13}}\tabularnewline
\hline 
Random Forest & 78.32 & 66.74 & 72.07 & 34.13 & 20.66 & 78.82 & 68.65 & 73.38 & 18.38 & 5.48\tabularnewline
\hline 
KNeighbors & 77.95 & \textbf{72.86} & 75.32 & 23.5 & 41.78 & \textcolor{red}{\textbf{80.27}} & 73.9 & 76.96 & 14.8 & 50.15\tabularnewline
\hline 
MLP & \textbf{79.90} & 71.36 & \textbf{75.39} & 1180.75 & 47.77 & 65.14 & 68.96 & 66.99 & 591.4 & 42.79\tabularnewline
\hline 
Naive Bayes & 65.73 & 51.77 & 57.92 & \textbf{22.77} & 4.52 & 43.47 & 59.67 & 50.3 & \textcolor{red}{\textbf{14.57}} & 1.08\tabularnewline
\hline 
\end{tabular}

\end{table*}

In terms of detection performance, it is shown from Tab.~\ref{tab:binary-4}, \ref{tab:binary-8} and \ref{tab:binary-16} that when the number of reduced features (i.e. extracted or selected) $K$ increases, the detection performance of feature extraction generally improves, while that of feature selection does not improve when we increase $K$ from 8 to 16. In fact, the precision, recall and F1-score of feature selection even slightly degrade from Tab.~\ref{tab:binary-8} to Tab.~\ref{tab:binary-16}. This phenomenon is understandable due to the fact that if the number of selected features gets larger, it is likely to have more noisy or unimportant features appearing in the selected ones, which are expected to deteriorate the detection performance. Moreover, comparing the two feature reduction methods, we find that when the number of reduced features is small, i.e., $K=4$, the detection performance of feature extraction is much better than that of feature selection. For instance, in Tab.~\ref{tab:binary-4}, the highest F1-score of feature extraction is 85.42\% when the KNeighbors classifier is used, while that of feature selection is lower with 81.94\% when the Decision Tree classifier is used. However, for larger $K$ such as 8 and 16 in Tab.~\ref{tab:binary-8} and  Tab.~\ref{tab:binary-16}, the feature selection method achieves better accuracy than its extraction counterpart, especially when using Decision Tree for classification. For example, when Decision Tree is employed in Tab.~\ref{tab:binary-8} to achieve the lowest inference time, the F1-score of feature selection is 87.47\%, which is higher than that of feature extraction with 85.69\%. It is also shown from Tab.~\ref{tab:binary-4}, \ref{tab:binary-8} and \ref{tab:binary-16} that when using feature selection, the Decision Tree classification method always provides the best precision, recall as well as F1-score. By contrast, the feature extraction method would enjoy the KNeighbors classifier when $K$ are small, i.e., 4 or 8, while Decision Tree is only its best classifier when $K$ becomes larger, such as $K=16$.

In terms of the runtime performance, Tabs.~\ref{tab:binary-4}, \ref{tab:binary-8} and \ref{tab:binary-16} demonstrate that both the training time and the inference time of feature selection is lower than that of feature extraction. This is because of the fact that the feature extraction method requires additional computational resources when compressing the high-dimensional data into low-dimensional data, as explained in Section~\ref{sec:idiot}, while the feature selection almost do not require any computing resources when just picking up $K$ out of $D$ features. More particularly, in Tab.~\ref{tab:binary-8}, the best inference time of feature selection is 0.11~$\mu$s, which is 36 times lower than that of feature extraction being 3.95~$\mu$s, where the Decision Tree classifier is the best choice for both feature reduction methods for minimizing the inference time. Again, Decision Tree is one of the best classifiers for minimizing both training and inference times, in addition to the Naive Bayes classifier, which however does not achieve a good accuracy. 

Finally, in order to better understand the attack detection performance of feature selection and feature extraction, in Tab.~\ref{tab:acc-binary-best}, we provide the accuracy comparison for each class in binary classification, namely, Normal and Abnormal. Similar to Tabs.~\ref{tab:binary-4}, \ref{tab:binary-8} and \ref{tab:binary-16}, we consider the number of reduced features $K$ being 4, 8 and 16. Besides, based on the results obtained from these three tables, we only include the classifiers that offer the highest F1-scores for accuracy comparison for each class in Tab.~\ref{tab:acc-binary-best}, namely, MLP and KNeighbors for feature extraction and Decision Tree for feature selection. Herein, the highest accuracy for each class with respect to $K$ is highlighted in bold, while the highest values in both feature selection and feature extraction are highlighted in bold and red color. It is worth noting from this table that in both feature reduction methods, while the accuracy of detecting Normal class steadily improves when increasing $K$, that of detecting Abnormal class gradually degrades. This interestingly indicates that in order to detect more attacks, we should select small $K$ rather than large $K$. In addition, Tab.~\ref{tab:acc-binary-best} shows that for both feature reduction methods, the accuracy of Abnormal class is much higher than that of Normal class. Observe the average accuracy from this table, we find that the accuracy of feature extraction is less sensitive to varying $K$ that that of feature selection, which varies significantly with respect to $K$.

\begin{table*}[ht]
\centering \caption{Feature Selection versus Feature Extraction: 16 selected/extracted features and multiclass classification}
\label{tab:multiclass-16} \linespread{1.0} %


\begin{tabular}{|l|c|c|c|c|c|c|c|c|c|c|}
\hline 
\multirow{2}{*}{Models} & \multicolumn{5}{c|}{Feature Extraction} & \multicolumn{5}{c|}{Feature Selection}\tabularnewline
\cline{2-11} \cline{3-11} \cline{4-11} \cline{5-11} \cline{6-11} \cline{7-11} \cline{8-11} \cline{9-11} \cline{10-11} \cline{11-11} 
 & P & R & F1 & training (s) & inference ($\mu$s) & P & R & F1 & training (s) & inference ($\mu$s)\tabularnewline
\hline 
Decision Tree & 77.33 & 70.11 & 73.55 & 37.68 & \textbf{5.04} & 79.59 & \textcolor{red}{\textbf{75.78}} & \textcolor{red}{\textbf{77.64}} & 15.17 & \textcolor{red}{\textbf{0.19}}\tabularnewline
\hline 
Random Forest & 78.36 & 66.71 & 72.07 & 53.12 & 21.27 & \textcolor{red}{\textbf{80.03}} & 68.02 & 73.54 & 22.17 & 5.69\tabularnewline
\hline 
KNeighbors & 77.56 & \textbf{72.03} & 74.69 & 34.94 & 1405.43 & 78.79 & 63.91 & 70.58 & 14.72 & 1396.85\tabularnewline
\hline 
MLP & \textbf{79.44} & 71.97 & \textbf{75.52} & 1428.97 & 47.51 & 64.42 & 69.01 & 66.63 & 895.37 & 49.51\tabularnewline
\hline 
Naive Bayes & 74.57 & 60.59 & 66.87 & \textbf{34.75} & 6.34 & 47.28 & 59.75 & 52.79 & \textcolor{red}{\textbf{14.47}} & 1.09\tabularnewline
\hline 
\end{tabular}

\end{table*}

\begin{table*}[ht]
\centering \caption{Accuracy comparison for each class between feature selection and feature extraction using multiclass classification}
\label{tab:acc-multiclass-best} \linespread{1.0} 
\begin{tabular}{|c|c|c|c|c|c|c|}
\hline 
\multirow{2}{*}{Class} & \multicolumn{3}{c|}{Feature Extraction (MLP)} & \multicolumn{3}{c|}{Feature Selection (Decision Tree)}\tabularnewline
\cline{2-7} \cline{3-7} \cline{4-7} \cline{5-7} \cline{6-7} \cline{7-7} 
 & $K=4$  & $K=8$  & $K=16$  & $K=4$ & $K=8$  & $K=16$ \tabularnewline
\hline 
Analysis & 0 & 0 & 0 & 1.03 & \textbf{\textcolor{red}{1.62}} & 0\tabularnewline
\hline 
Backdoor & 0 & 0 & 0 & 6 & \textbf{\textcolor{red}{7.2}} & 6.17\tabularnewline
\hline 
DoS & 0.81 & 10.61 & \textbf{11.2} & 8.12 & \textbf{\textcolor{red}{14.18}} & 12.96\tabularnewline
\hline 
Exploits & \textbf{\textcolor{red}{86.79}} & 85.92 & 85.59 & 53.33 & \textbf{84.65} & 82.79\tabularnewline
\hline 
Fuzzers & \textbf{\textcolor{red}{67.6}} & 66.46 & 58 & 43.1 & \textbf{50.46} & 49.64\tabularnewline
\hline 
Generic & 96.22 & 96.24 & \textbf{96.3} & \textbf{\textcolor{red}{98.17}} & 98.04 & 97.38\tabularnewline
\hline 
Normal & 62.47 & 62.5 & \textbf{68.79} & 59.19 & \textbf{\textcolor{red}{77.06}} & 76.51\tabularnewline
\hline 
Reconnaissance & 50.69 & 60.55 & \textbf{66.5} & 39.99 & \textbf{\textcolor{red}{78.58}} & 78.4\tabularnewline
\hline 
Shellcode & 0 & 7.67 & \textbf{15.08} & 0 & \textbf{\textcolor{red}{47.62}} & 42.86\tabularnewline
\hline 
Worms & 0 & 2.27 & \textbf{9.09} & 11.36 & \textbf{\textcolor{red}{61.36}} & 34.09\tabularnewline
\hline 
Average & 69.03 & 69.79 & \textbf{72.29} & 61.65 & \textbf{\textcolor{red}{76.62}} & 75.78\tabularnewline
\hline 
\end{tabular}

\end{table*}

\begin{table*}[ht]
\centering \caption{Accuracy comparison for each class between feature selection and feature extraction using multiclass classification and the same Decision Tree classifier}
\label{tab:acc-multiclass-DT} \linespread{1.0} 

\begin{tabular}{|c|c|c|c|c|c|c|}
\hline 
\multirow{2}{*}{Class} & \multicolumn{3}{c|}{Feature Extraction (Decision Tree)} & \multicolumn{3}{c|}{Feature Selection (Decision Tree)}\tabularnewline
\cline{2-7} \cline{3-7} \cline{4-7} \cline{5-7} \cline{6-7} \cline{7-7} 
 & $K = 4$ & $K = 8$ & $K = 16$ & $K = 4$ & $K = 8$ & $K = 16$\tabularnewline
\hline 
Analysis & 10.78 & \textbf{\textcolor{red}{11.82}} & 11.52 & 1.03 & \textbf{1.62} & 0\tabularnewline
\hline 
Backdoor & 2.92 & 2.74 & \textbf{5.83} & 6 & \textbf{\textcolor{red}{7.2}} & 6.17\tabularnewline
\hline 
DoS & 19.93 & 19.81 & \textbf{
\textcolor{red}{24.75}} & 8.12 & \textbf{14.18} & 12.96\tabularnewline
\hline 
Exploits & 59.07 & 61.77 & \textbf{63.94} & 53.33 & \textbf{\textcolor{red}{84.65}} & 82.79\tabularnewline
\hline 
Fuzzers & 38.16 & 42.08 & \textbf{44.46} & 43.1 & \textbf{\textcolor{red}{50.46}} & 49.64\tabularnewline
\hline 
Generic & 93.58 & \textbf{96.22} & 95.82 & \textbf{\textcolor{red}{98.17}} & 98.04 & 97.38\tabularnewline
\hline 
Normal & 72.48 & \textbf{73.13} & 72.92 & 59.19 & \textbf{\textcolor{red}{77.06}} & 76.51\tabularnewline
\hline 
Reconnaissance & 36.87 & 42.68 & \textbf{45.68} & 39.99 & \textbf{\textcolor{red}{78.58}} & 78.4\tabularnewline
\hline 
Shellcode & 23.54 & 28.04 & \textbf{32.8} & 0 & \textbf{\textcolor{red}{47.62}} & 42.86\tabularnewline
\hline 
Worms & \textbf{15.91} & 13.64 & 11.36 & 11.36 & \textbf{\textcolor{red}{61.36}} & 34.09\tabularnewline
\hline 
Average & 67.6 & 69.42 & \textbf{70.11} & 61.65 & \textbf{\textcolor{red}{76.62}} & 75.78\tabularnewline
\hline 
\end{tabular}

\end{table*}

\begin{table*}[ht]
\centering \caption{Accuracy comparison for each class between feature selection and feature extraction using multiclass classification and the same MLP classifier}
\label{tab:acc-multiclass-MLP} \linespread{1.0} 
\begin{tabular}{|c|c|c|c|c|c|c|}
\hline 
\multirow{2}{*}{Class} & \multicolumn{3}{c|}{Feature Extraction (MLP)} & \multicolumn{3}{c|}{Feature Selection (MLP)}\tabularnewline
\cline{2-7} \cline{3-7} \cline{4-7} \cline{5-7} \cline{6-7} \cline{7-7} 
 & $K=4$  & $K=8$  & $K=16$  & $K=4$ & $K=8$  & $K=16$ \tabularnewline
\hline 
Analysis & 0 & 0 & 0 & 0 & 0 & 0\tabularnewline
\hline 
Backdoor & 0 & 0 & 0 & 0 & 0 & 0\tabularnewline
\hline 
DoS & 0.81 & 10.61 & \textbf{\textcolor{red}{11.2}} & \textbf{6.6} & 0.83 & 0.42\tabularnewline
\hline 
Exploits & \textbf{\textcolor{red}{86.79}} & 85.92 & 85.59 & 19.16 & 25 & \textbf{26.23}\tabularnewline
\hline 
Fuzzers & \textbf{\textcolor{red}{67.6}} & 66.46 & 58 & \textbf{30.86} & 18.51 & 11.86\tabularnewline
\hline 
Generic & 96.22 & 96.24 & \textbf{96.3} & \textbf{\textcolor{red}{96.96}} & 96.2 & 96.21\tabularnewline
\hline 
Normal & 62.47 & 62.5 & \textbf{68.79} & 68.66 & 75.5 & \textbf{\textcolor{red}{93.08}}\tabularnewline
\hline 
Reconnainssance & 50.69 & 60.55 & \textbf{66.5} & 0.11 & \textbf{\textcolor{red}{73.31}} & 36.61\tabularnewline
\hline 
Shellcode & 0 & 7.67 & \textbf{\textcolor{red}{15.08}} & 0 & \textbf{1.06} & 0\tabularnewline
\hline 
Worms & 0 & 2.27 & \textbf{\textcolor{red}{9.09}} & 0 & 0 & 0\tabularnewline
\hline 
Average & 69.03 & 69.79 & \textbf{\textcolor{red}{72.29}} & 58.38 & 63.88 & \textbf{69.88}\tabularnewline
\hline 
\end{tabular}

\end{table*}

\begin{table*}[ht]
\centering \caption{A summary of comparison between feature extraction and feature selection}
\label{tab:sum-comparison} \linespread{1.0} 
\begin{tabular}{|c|l|c|c|}
\hline 
No. & Content & Extraction & Selection\tabularnewline
\hline 
\hline 
1 & Higher accuracy when $K$ is very small, such as $K=4$ & \checkmark  & \tabularnewline
\hline 
2 & Higher accuracy when $K$ gets large, such as $K=8$ or 16 &  & \checkmark \tabularnewline
\hline 
3 & Lower training time &  & \checkmark \tabularnewline
\hline 
4 & Lower inference time &  & \checkmark \tabularnewline
\hline 
5 & Detect more diverse attack types when using the same classifier & \checkmark  & \tabularnewline
\hline 
6 & Less sensitive to the number of selected/extracted features $K$ & \checkmark  & \tabularnewline
\hline 
7 & MLP is the best classifier & \checkmark  & \tabularnewline
\hline 
8 & Decision Tree is the best classifier &  & \checkmark \tabularnewline
\hline 
9 & Detection performance degrades when $K$ is too large &  & \checkmark \tabularnewline
\hline 
10 & Detection performance steadily improves when $K$ increases & \checkmark  & \tabularnewline
\hline 
12 & Higher accuracy in detecting Exploits and Fuzzers classes & \checkmark  & \tabularnewline
\hline 
13 & Higher accuracy in detecting 8 remaining classes, except for Exploits and Fuzzers  &  & \checkmark\tabularnewline
\hline 
14 & Accuracy of Abnormal class is much higher than that of Normal class (Binary) & \checkmark & \checkmark\tabularnewline
\hline 
15 & Accuracy of detecting Abnormal class degrades when $K$ increases
(Binary) & \checkmark & \checkmark\tabularnewline
\hline 
16 & Higher accuracy for Exploits, Generic, Normal, Reconnaissance than
remaining classes & \checkmark  &  
 \checkmark \tabularnewline
\hline 
\end{tabular}

\end{table*}





\subsection{Multiclass classification}
\label{subsec:multiclass} 
\vspace{-0.1cm}
We compare both the detection performance and runtime of feature selection and feature extraction in Tabs.~\ref{tab:multiclass-4}, \ref{tab:multiclass-8}, and \ref{tab:multiclass-16} for 4, 8, and 16 selected/extracted features, respectively, when multiclass classification is considered. Here, we still employ five machine learning models as in binary classification. As shown via these three tables,  similar to the binary case, the precision, recall and F1-score of both methods generally improve when increasing the number of reduced features $K$. For example, the highest F1-scores of feature extraction are 74.11\%, 75.39\%, and 75.52\%, while that of feature selection are 65.43\%, 78.36\% and 77.64\%, when $K$ = 4, 8, and 16 reduced features, respectively. As such, feature extraction outperforms its counterpart when $K$ is small such as $K=4$, however, this is no longer true when $K$ gets larger such as $K=8$ and $16$, where feature selection performs much better than feature extraction. Again, akin to the binary classification, it is shown from Tabs.~\ref{tab:multiclass-8} and \ref{tab:multiclass-16} that the detection performance of feature selection degrades when $K$ increases from 8 to 16, mostly due to the impact of noisy or irrelevant features when having more features selected. 

Besides, unlike the binary case, where KNeighbors is the best classifier for feature extraction when $K$ is small such as 4 and 8, with multiclass classification, MLP now provides the best detection performance of feature extraction for any values of $K$, as shown via Tabs.~\ref{tab:multiclass-4}, \ref{tab:multiclass-8}, and \ref{tab:multiclass-16}. Meanwhile, feature selection still enjoys the Decision Tree classifier to achieve the highest detection performance, similar to the binary classification analyzed in the previous subsection, while MLP does not offer a good detection performance for feature selection. Additionally, the Naive Bayes classifier achieves the worst accuracy for both feature reduction methods.

With regard to the runtime comparison, again, Tabs.~\ref{tab:multiclass-4}, \ref{tab:multiclass-8}, and \ref{tab:multiclass-16} demonstrate that  the training and inference times of feature selection are significantly  lower  than that of feature extraction. For example, using the same Decision Tree model for achieving the lowest runtime, in Tab.~\ref{tab:multiclass-8} when $K=8$, the inference time of feature selection is 0.19~$\mu$s, which is 26 times lower than that of feature extraction with 5.04~$\mu$s. Similarly, it is shown from this table that the training time of feature selection is also 2 times lower than that of its extraction counterpart. In addition, the Decision Tree model  provides the lowest inference time for both feature reduction methods, while the neural network-based MLP classifier exhibits both the highest inference and training times for them.

Finally, we compare the accuracy for detecting each attack type (including 9 attack classes and 1 normal class, as described in Section~\ref{sec:dataset}) between feature selection and feature extraction in Tab.~\ref{tab:acc-multiclass-best}, where the values of $K$ are 4, 8, and 16 reduced features. Herein, we employ MLP and Decision Tree classifiers for feature extraction and feature selection, respectively, in order to achieve the best detection performance, as analyzed in the previous discussions. It is observed from Tab.~\ref{tab:acc-multiclass-best} that feature selection performs better than feature extraction in most of classes, except for Exploits and Fuzzers classes. This table also shows that both methods are capable of achieving higher accuracy for Exploits, Generic, Normal and Reconnaissance classes than the remaining ones. Additionally, similar to the binary classification discussed in Tab.~\ref{tab:acc-binary-best}, the multiclass classification accuracy of feature extraction is less sensitive to the number of reduced features $K$ than that of its selection counterpart. More importantly, feature selection with MLP is unable to correctly detect any samples of Analysis and Backdoor, even for all three values of $K$. By contrast, feature selection with Decision Tree classifier is capable of correctly detecting samples from all classes. We found that this is mainly due to the machine learning classifier rather than the feature reduction method we choose. In order to clarify this issue, we compare the accuracy for each class between the two feature reduction methods using the same Decision Tree and MLP classifiers in Tab.~\ref{tab:acc-multiclass-DT} and Tab.~\ref{tab:acc-multiclass-MLP}, respectively. It is shown via Tab.~\ref{tab:acc-multiclass-MLP} that using the same MLP, similar to feature extraction, feature selection is unable to detect any samples of Analysis and Backdoor correctly. Observe these two tables, we found that if the same classifier is employed, feature extraction tends to be able to detect more diverse attack types than feature selection. This is due to the fact that feature extraction can extract key information from all available features, leading to more diverse attack types, instead of relying solely on a subset of selected features as in the feature selection approach. In other words, feature selection tends to detect only attack types, which are highly correlated to the features it selects.

\textcolor{black}{In summary, considering both binary and multiclass classification for the NIDS, the feature selection method not only provides better detection performance but also lower training and inference time compared to its feature extraction counterpart, especially when the number of reduced features $K$ increases. However, the feature extraction method is much more reliable than its selection counterpart, particularly when $K$ is very small, such as $K=4$. Additionally, among five considered classifiers, while Decision Tree is the best classifier for improving the accuracy of feature selection, a neural network-based MLP is the best one for feature extraction. Last but not least, feature extraction is less sensitive to changing the number of reduced features $K$ than feature selection, and this holds true for both binary and multiclass classifications. For more details, we provide a comprehensive comparison between feature selection and feature extraction in intrusion detection systems in Tab.~\ref{tab:sum-comparison}.}





\section{Conclusions\label{sec:Conclusions}}
 
 \textcolor{black}{We have compared two typical machine learning-based intrusion detection methods, namely, feature selection and feature extraction, in the presence of the modern UNSW-NB15 dataset, where both binary and multiclass classifications were considered.} \textcolor{black}{Our extensive comparison showed that when the number of reduced features is large enough, such as 8 or 16, feature selection not only achieves higher detection accuracy, but also requires less  training and inference times than feature extraction. However, when the number of reduced features is very small, such as 4 or less, feature extraction notably outperforms its selection counterpart. Besides, the detection performance of feature selection tends to degrade when the number of selected features becomes too large, while that of feature extraction steadily improves. We also found that while MLP is the best classifier for feature extraction, Decision Tree is the best one for feature selection for achieving the highest attack detection accuracy. Finally, our accuracy analysis for each attack class demonstrated that feature extraction is not only less sensitive to varying the number of reduced features but also capable of detecting more diverse attack types  than feature selection. Both tend to be able to detect more attacks, i.e., Abnormal classes, when having more features selected or extracted. We believe that such insightful observations about the performance comparison between two feature reduction methods give us a helpful guideline on choosing a suitable intrusion detection method for each specific scenario.} \textcolor{black}{Finally, note that our study evaluated the effectiveness of feature reduction methods only on the UNSW-NB15 dataset. In the future, we intend to explore whether our observations with UNSW-NB15 are applicable to other intrusion detection datasets, such as, NSL-KDD, KDD99, CICIDS2017, and DARPA1998. We also plan to thoroughly investigate the performance of various deep learning classification models for NIDS, and compare with existing machine learning models.} \\






\textbf{Declarations} 

\textbf{Author Contribution} Vu-Duc Ngo and Tuan-Cuong Vuong wrote the main manuscript. Thien Van Luong and Hung Tran reviewed and corrected the manuscript.

\textbf{Ethical Approval} Not applicable.

\textbf{Acknowledgement} This work was supported in part by the SSF Framework Grant Serendipity and R\&D project of Brighter Gates AB, Sweden. 

\textbf{Data Availability} The paper does not include any supporting data.

\textbf{Conflicts of interests} All authors declare that they do not have any conflict of interest.

\textbf{Funding} Not applicable.

\bibliographystyle{IEEEtran}
\bibliography{IEEEabrv,refs}

\begin{IEEEbiography}[{{\includegraphics[clip,width=1in,height=1.25in]{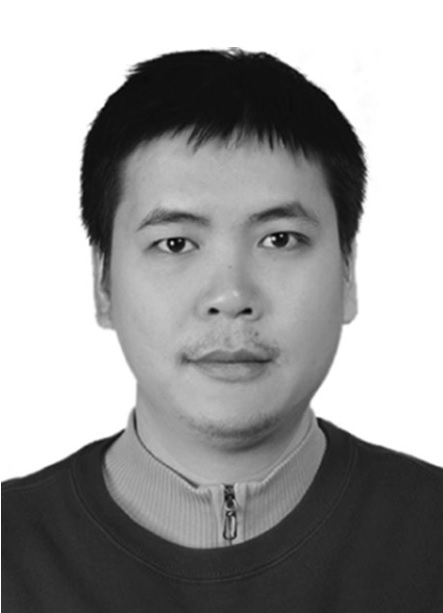}}}]{Vu-Duc Ngo}
received the Ph.D. degree from the Korea Advanced Institute of Science and Technology in 2011. From 2007 to 2009, he was a Co-Founder and the CTO of Wichip Technologies Inc., USA. Since 2009, he has been a Co-Founder and the Director of uVision Jsc, Vietnam. Since 2012, he has been serving as a BoM Member of the National Program on Research, Training, and Construction, High-Tech Engineering Infrastructure of Vietnam. He is currently a Researcher with the \mbox{MobiFone} Research and Development Center, MobiFone Corporation, and also a Lecturer with the School of Electrical and Electronics Engineering, Hanoi University of Science and Technology, Hanoi, Vietnam. His research interests are in the fields of SoC, NoC design and verification, VLSI design for multimedia codecs, and wireless communications PHY layer. He was a recipient of the IEEE 2006 ICCES, the IEEE 2012 ATC and the NICS 2021 Best Paper Awards.
\end{IEEEbiography}

\begin{IEEEbiography}[{{\includegraphics[clip,width=1in,height=1.25in]{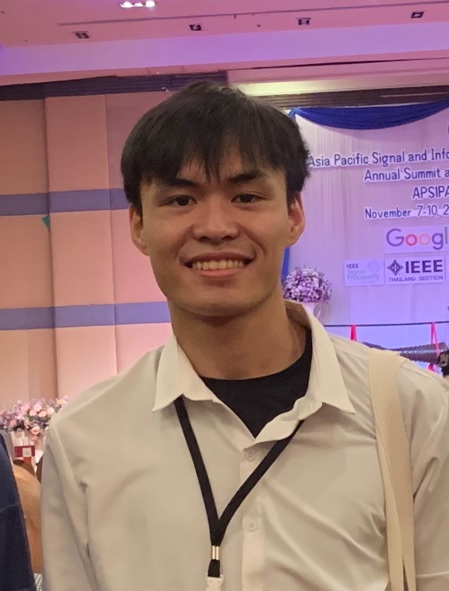}}}]{Tuan-Cuong Vuong}
is currently a second year Bachelor student, working at AIoT Lab, Faculty of Computer Science, Phenikaa University, Hanoi, Vietnam. His research interests include applied machine learning and deep learning in cyber security and computer vision.
\end{IEEEbiography}

\begin{IEEEbiography}[{{\includegraphics[clip,width=1in,height=1.25in]{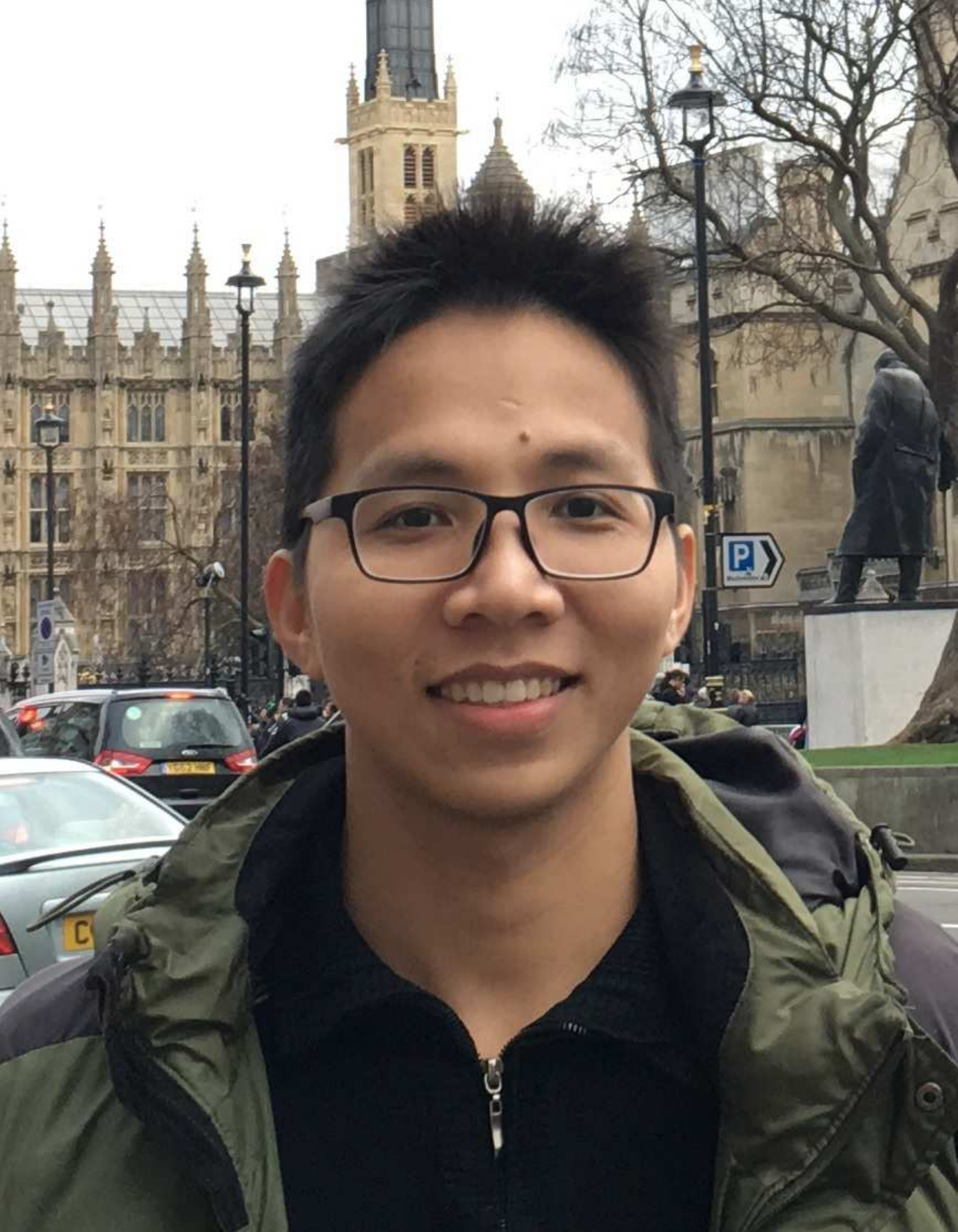}}}]{Thien Van Luong}
is currently a Lecturer with the Faculty of Computer Science, and a Leader of AIoT Lab (https://aiot.phenikaa-uni.edu.vn/), Phenikaa University, Vietnam. He was a Research Fellow with University of Southampton, U.K. Prior to that he obtained the Ph.D. degree at Queen's University Belfast, U.K., and the B.S. degree at Hanoi University of Science and Technology, Vietnam. His research interests include applied machine learning in signal processing and wireless communications.
\end{IEEEbiography}

\begin{IEEEbiography}[{{\includegraphics[clip,width=1in,height=1.25in]{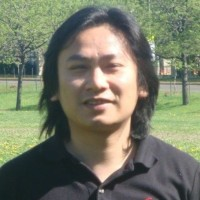}}}]{Hung Tran}
Hung Tran received the B.S. and M.S. degrees in information technology from Vietnam National University, Hanoi, Vietnam, in 2002 and 2006, respectively, and the Ph.D. degree from the Blekinge Institute of Technology, Sweden, in March 2013. In 2014, he was with the Electrical Engineering Department, ETS, Montreal, Canada. From 2015 to 2020, he was a Researcher at Malardalen University, Sweden. He is currently working as a Lecturer at the Computer Science Department, Phenikaa University, Vietnam. Besides doing research in the areas of wireless communication, he is also interested in topics of natural language processing and artificial intelligence, which have been applied to develop core engines for the academic gates platform (https://www.academicgates.com).
\end{IEEEbiography}

\end{document}